\documentclass[prl,aps,twocolumn,showpacs]{revtex4}
\usepackage{epsfig}
\usepackage{bm}

\begin{document}

\title{Transitional solar dynamics, cosmic rays and global warming}

\author{\small  A. Bershadskii}
\affiliation{\small {ICAR, P.O. Box 31155, Jerusalem 91000, Israel }}

\begin{abstract}
Solar activity is studied using a cluster analysis of the time-fluctuations of the sunspot
number. It is shown that in an Historic period 
the high activity components of the solar cycles exhibit strong clustering,
whereas in a Modern period (last seven solar cycles: 1933-2007) they exhibit a 
white-noise (non-)clustering behavior. Using this observation it is shown
that in the Historic period, emergence of the sunspots in the solar
photosphere was strongly dominated by turbulent photospheric
convection. In the Modern period, this domination was broken by a
new more active dynamics of the inner layers of the convection zone.
Then, it is shown that the dramatic change of the sun dynamics at
the transitional period (between the Historic and Modern periods,
solar cycle 1933-1944yy) had a clear detectable impact on Earth 
climate. A scenario of a chain of 
transitions in the solar convective zone is suggested in order to explain 
the observations, and a forecast for the global warming is suggested on the 
basis of this scenario. A relation between the recent transitions and solar long-period {\it chaotic} 
dynamics has been found. Contribution of the galactic turbulence (due to galactic cosmic rays)
has been discussed. These results are also considered in a content of {\it chaotic} climate dynamics at
millennial timescales.

\end{abstract}


\maketitle

\section{Introduction}

\begin{figure} \vspace{-0.5cm}\centering
\epsfig{width=.45\textwidth,file=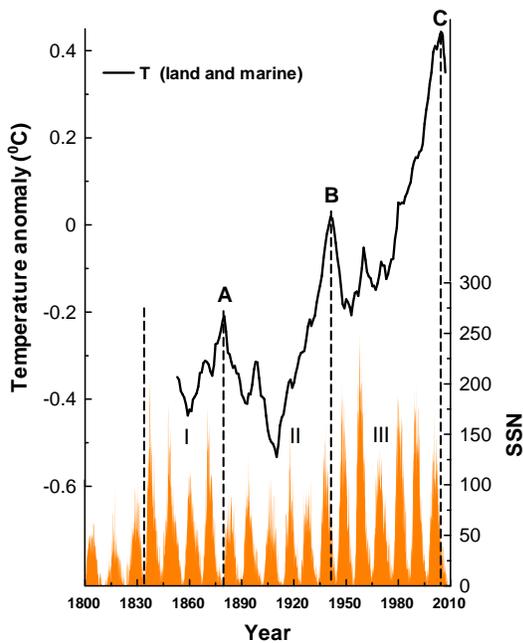} \vspace{-2.3cm}
\caption{Sunspot number (SSN, monthly) vs time \cite{belg}. The dashed
straight lines separate between periods of different intensity of the 
solar activity. The solid curve shows the global temperature anomaly (combined land and
marine, 7-years running average) \cite{temp}.}
\end{figure}
 The sunspot number is the main direct and
reliable source of information about the solar dynamics for historic
period. This information is crucial, for instance, for analysis of a
possible connection between the sun activity and the global warming.
In a recent papers \cite{sol},\cite{usos1} results of a
reconstruction of the sunspot number were presented for the past
11,400 years. The reconstruction shows that \cite{sol}: "..the level of solar activity during the past 
70 years is exceptional, and the previous period of equally high activity occurred more than 8,000 years ago" 
(section III in Fig. 1) . 

Figure 1 gives first indication ('by eye'), that the prominent maxima of the global temperature data 
(solid curve in the Fig. 1) correspond to transitions between periods 
of different intensity of the Sun activity (characterized by the monthly sunspot number-SSN). 
This observation can be considered as an indication of a strong impact of the solar activity {\it transitions} 
on the global Earth climate. Therefore, understanding of the physical processes in the Sun, 
which cause these activity transitions, seems to be crucial for any serious forecast for 
global Earth climate.

\section{Sunspots}

When magnetic field lines are twisted and poke through the solar
photosphere the sunspots appear as the visible counterparts of
magnetic flux tubes in the convective zone of the sun. Since a
strong magnetic field is considered as a primary phenomenon that
controls generation of the sunspots the crucial question is: Where
has the magnetic field itself been generated? The location of the
solar dynamos is the subject of vigorous discussions in recent years. A
general consensus had been developed to consider
the shear layer at the {\it bottom} of the convection zone as the main
source of the solar magnetic field \cite{sw} (see, for a recent
review \cite{brand1}). In recent years, however, the existence of a
prominent radial shear layer near the top of the convection zone has
become rather obvious and the problem again became actual. The presence
of large-scale meandering flow fields (like jet streams), banded
zonal flows and evolving meridional circulations together with
intensive multiscale turbulence shows that the near surface layer is
a very complex system, which can significantly affect the processes
of the magnetic field and the sunspots generation. There could be 
two sources for the poloidal magnetic field: one near the
bottom of the convection zone (or just below it \cite{sw}), another 
resulting from an active-region tilt near the surface of the
convection zone. For the recently renewed Babcock-Leighton
\cite{bab},\cite{leig} solar dynamo scenario, for instance, a
combination of the sources was assumed for predicting future solar
activity levels \cite{dtg}, \cite{ccj}. In this scenario the surface
generated poloidal magnetic field is carried to the bottom of the
convection zone by turbulent diffusion or by the meridional
circulation. The toroidal magnetic field is produced from this poloidal field 
by differential rotation in the bottom shear layer.
Destabilization and emergence of the toroidal fields (in the form of
curved tubes) due to magnetic buoyancy can be considered as a source
of pairs of sunspots of opposite polarity. The turbulent
convection in the convection zone and, especially, in the
near-surface layer captures the magnetic flux tubes and either {\it
disperses} or {\it pulls} them trough the surface to become
sunspots.

The magnetic field plays a passive role in the photosphere
and does not participate significantly in the turbulent photospheric 
energy transfer. On the other hand, the very complex and turbulent
near-surface layer (including photosphere) can significantly affect
the process of emergence of sunspots. The 
similarity of light elements properties in the spot umbra and
granulation is one of the indication of such phenomena.

The present paper reports a direct relation between the
fluctuations in sunspot number and the temperature of photospheric turbulent convection 
in the Historic period. 
This relation allows for certain conclusions
about the generation mechanisms of the magnetic fields and the
sunspots. In the Modern
period (section III in Fig. 1, cf. \cite{sol},\cite{usos1}) 
the relative role of the surface layer
(photosphere) in the process of emergence of sunspots decreased
in comparison with the Historic period (section II in Fig. 1), 
implying a drastic increase of the relative role of the inner layers 
of the convection zone in the Modern period.

\section{Time-clustering of fluctuations} 

In order to extract new
information from sunspot number data we apply the
fluctuation clustering analysis suggested in the Ref. \cite{sb} (see also Ref. \cite{bEPL}). For 
a time depending signal we count the number of
'zero'-crossings of the signal (the points on the time axis
where the signal is equal to zero) in a time interval $\tau$ and
consider their running density $n_{\tau}$. Let us denote
fluctuations of the running density as $\delta n_{\tau} = n_{\tau} -
\langle n_{\tau} \rangle$, where the brackets mean the average over
long times. We are interested in scaling variation of the standard
deviation of the running density fluctuations $\langle \delta
n_{\tau}^2 \rangle^{1/2}$ with $\tau$
$$
\langle \delta n_{\tau}^2 \rangle^{1/2} \sim \tau^{-\alpha}
\eqno{(1)}
$$
For white noise signal it can be derived analytically
\cite{molchan},\cite{lg} that $\alpha = 1/2$ (see also \cite{sb}).
The same consideration can be applied not only to the
'zero'-crossing points but also to any level-crossing points of the
signal.
\begin{figure} \vspace{-0.5cm}\centering
\epsfig{width=.45\textwidth,file=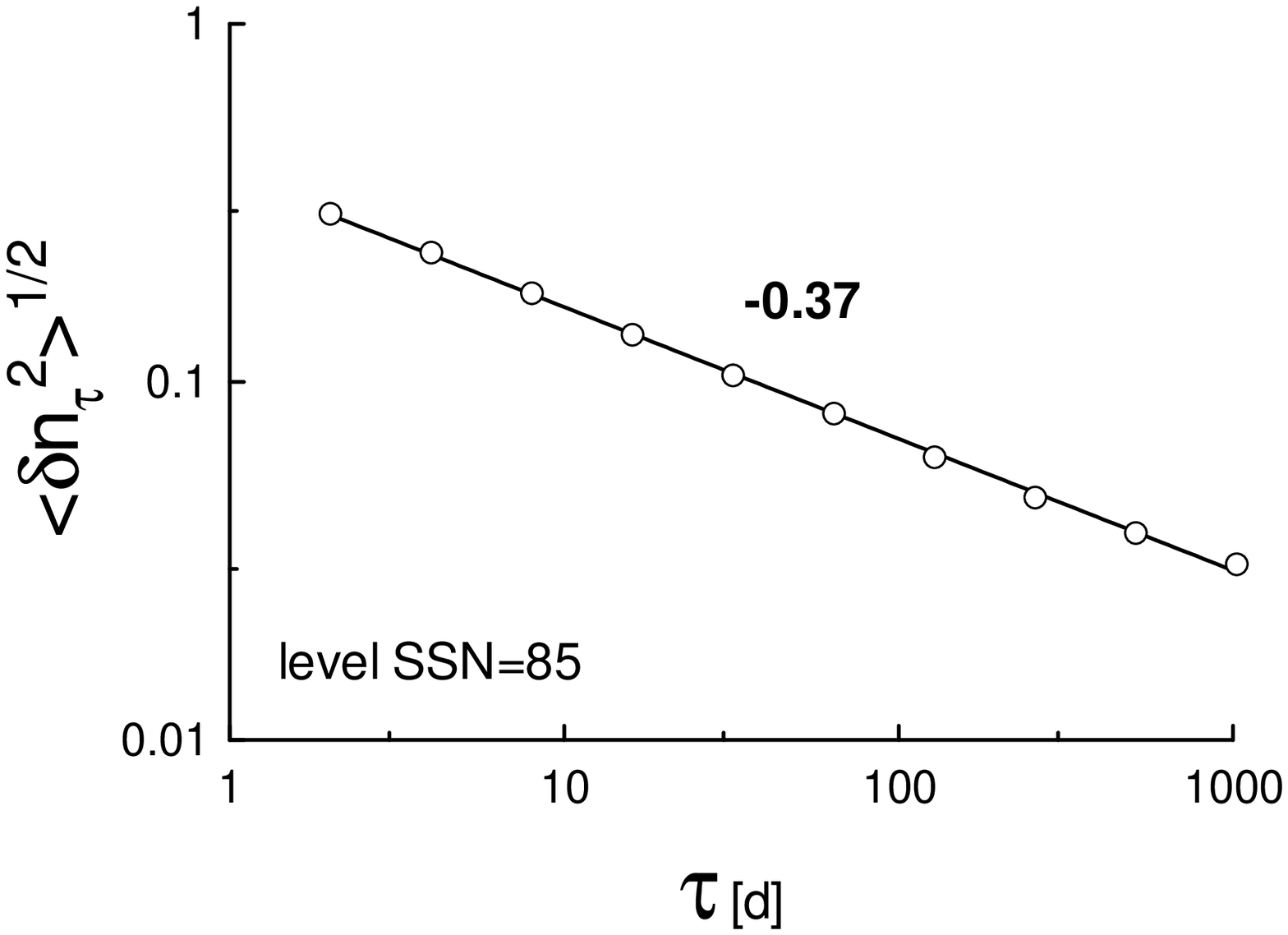} \vspace{-4cm}
\caption{The standard deviation for $\delta n_{\tau}$ vs $\tau$ 
(Historic period) in log-log scales. The straight
line (the best fit) indicates the scaling law Eq. (1). }
\end{figure}

\begin{figure} \vspace{-0cm}\centering
\epsfig{width=.45\textwidth,file=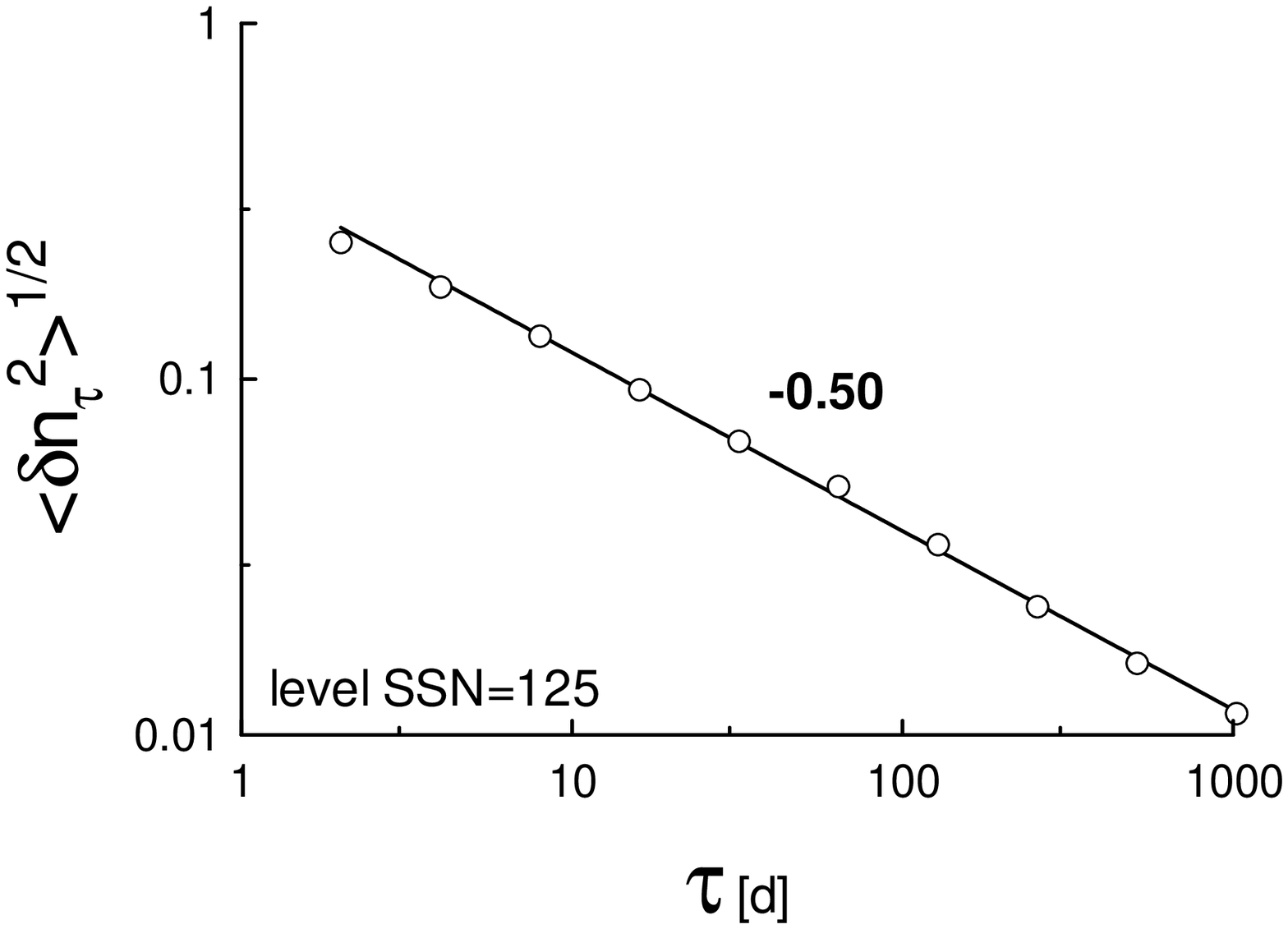} \vspace{-4.5cm}
\caption{The standard deviation for $\delta n_{\tau}$ vs $\tau$ 
(Modern period) in log-log scales. The straight
line (the best fit) indicates the scaling law Eq. (1).  }
\end{figure}
\begin{figure} \vspace{-0.3cm}\centering
\epsfig{width=.45\textwidth,file=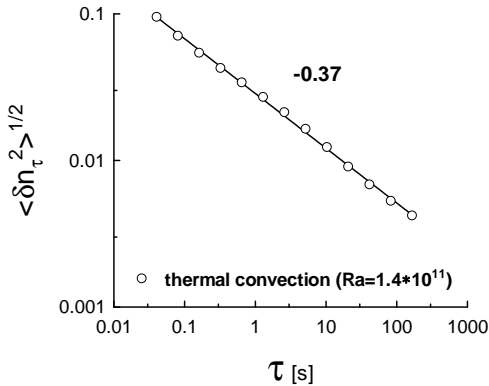} \vspace{-4.2cm}
\caption{The standard deviation for $\delta n_{\tau}$ vs $\tau$ for
the temperature fluctuations in the Rayleigh-Bernard convection
laboratory experiment with $Ra=1.4 \times 10^{11}$ \cite{ns}. The
straight line indicates the scaling law Eq. (1). }
\end{figure}

  One can see that for the Modern
period (section III in Fig. 1) the solar activity 
is significantly different from that for the Historic period 
(section II in Fig. 1). Therefore, in order to
calculate the cluster exponent (if exists) for this signal one
should make this calculation separately for the Modern and for the
Historic periods. We are interested in the active parts of the solar
cycles. Therefore, for the Historic period let us start from the
level SSN=85. The set of the level-crossing points has a few large
voids corresponding to the weak activity periods. In the telegraph signal Ref. [10], 
corresponding to the data set, the large voids became so clear detectable that there 
is no problem to cut them off. Then, the remaining data have been merged providing 
a statistically stationary set (about $10^4$ data points). The robustness of the 
procedure has been successfully checked. 

Fig. 2 shows (in the log-log scales) dependence of the standard deviation of the running
density fluctuations $\langle \delta n_{\tau}^2 \rangle^{1/2}$ on
$\tau$ for this data set. The straight line is drawn in this figure to indicate the
scaling (1). The slope of this straight line provides us with the
cluster-exponent  $\alpha = 0.37 \pm 0.02$. This value turned out 
to be insensitive to a reasonable variation of the SSN level. Results of
analogous calculations (including the large voids 'cut off' procedure) performed for 
the Modern period are shown in
Fig. 3 for the SSN level SSN=125. The calculations performed for the
Modern period provide us with the cluster-exponent $\alpha=0.5 \pm
0.02$ (and again this value turned out to be insensitive to a reasonable
variation of the SSN level).
\begin{figure} \vspace{-0.5cm}\centering
\epsfig{width=.45\textwidth,file=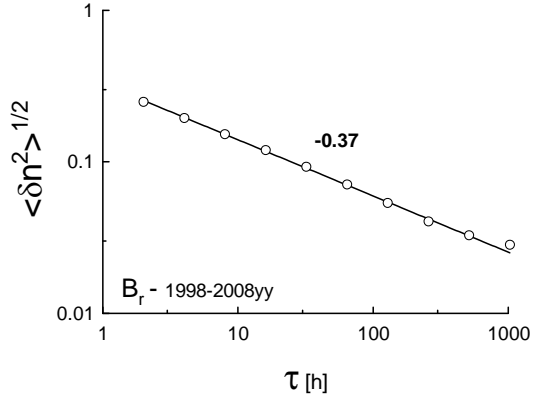} \vspace{-5cm} \caption{
The standard deviation for $\delta n_{\tau}$ vs. $\tau$ 
for radial component $B_r$ 
of the interplanetary magnetic field, as measured by
the ACE magnetometers for the last solar cycle (hourly average \cite{ACE}). 
}
\end{figure}

The exponent $\alpha\simeq 0.5$ (for the Modern period) indicates a
random (white noise like) situation. While the exponent
$\alpha\simeq 0.37$ (for the Historic period) indicates strong
clustering. The question is: Where is this strong clustering coming
from? It is shown in the paper \cite{sb} that turbulence produces 
signals with strong clustering. Moreover, the cluster
exponents for these signals depend on the turbulence intensity and
they are nonsensitive to the types of the boundary conditions.
Fortunately, we have direct estimates of the value of the main
parameter characterizing intensity of the turbulent convection in
photosphere: Rayleigh number $Ra \sim 10^{11}$ (see, for instance
\cite{bld}). In Fig. 4 we show calculation of the cluster exponent
for the temperature fluctuations in the classic Rayleigh-Bernard
convection laboratory experiment for $Ra \sim 10^{11}$ (for a
description of the experiment details see \cite{ns}). The calculated
value of the cluster exponent $\alpha =0.37\pm 0.01$ coincides with
the value of the cluster-exponent obtained above for the sunspot
number fluctuations for the Historic period. The value of the
Rayleigh number $Ra $ in the photosphere for the Historic period has the 
same order as for the Modern period: $Ra \sim10^{11}$ (see next Section). 
Therefore, the photospheric temperature fluctuations can produce the strong 
clustering of the sunspot number fluctuations for the Historic period. This
seems to be natural for the case when the photospheric convection
{\it determines} the sunspot emergence in the photosphere. However,
in the case when the effect of the photospheric convection on the
SSN fluctuations is comparable with the effects of the inner
convection zone layers on the SSN fluctuations the clustering should
be randomized by the mixing of the sources, and the cluster exponent
$\alpha \simeq 0.5$ (similar to the white noise signal). The last
case apparently takes place for the modern period.

Since the Rayleigh number $Ra$ of the photospheric convection preserves 
its order $Ra \sim 10^{11}$ with transition from
the Historic period to the Modern one (see next Section), we can assume that just
significant changes of the dynamics of the inner layers of the
convection zone (most probably - of the bottom layer) were the main reasons 
for the transition from the Historic to the Modern period.

\section{Magnetic fields in the solar wind and on Earth}

Although for the Modern period the turbulent convection in the photosphere has no 
decisive impact on the sunspots {\it emergence}, the large-scale properties of the 
magnetic field coming through the sunspots into the photosphere and then to the 
interplanetary space (so-called solar wind) can be strongly affected by the photospheric motion. 
In order to be detected the characteristic {\it scaling} scales of this impact should be larger than 
the scaling scales of the interplanetary turbulence (cf. \cite{b1},\cite{gold}). In particular, 
one can expect that the cluster-exponent of the large-scale interplanetary magnetic 
field (if exists) should be close to $\alpha \simeq 0.37$. In Fig. 5 we show cluster-exponent 
of the large-scale fluctuations of the radial component $B_r$ of the interplanetary 
magnetic field. For computing this exponent, we have used the hourly averaged data obtained from
Advanced Composition Explorer (ACE) satellite magnetometers for the last solar cycle \cite{ACE}. 
As it was expected the cluster-exponent $\alpha \simeq 0.37 \pm 0.02$. Analogous result was 
obtained for other components of the interplanetary magnetic field as well. 
\begin{figure} \vspace{-1cm}\centering
\epsfig{width=.45\textwidth,file=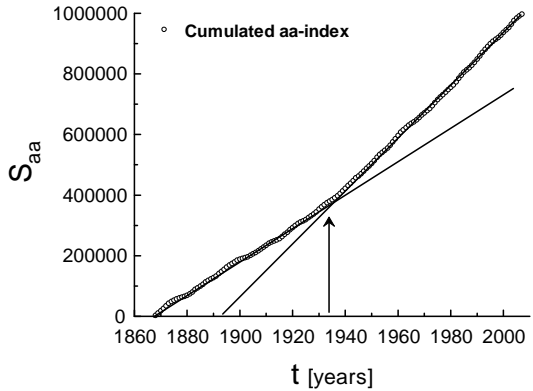} \vspace{-4.5cm} \caption{
Cumulative aa-index vs. time. Daily aa-index was taken from \cite{wdc}. 
The arrow indicates beginning of the transitional solar cycle.}
\end{figure}
\begin{figure} \vspace{-0.5cm}\centering
\epsfig{width=.45\textwidth,file=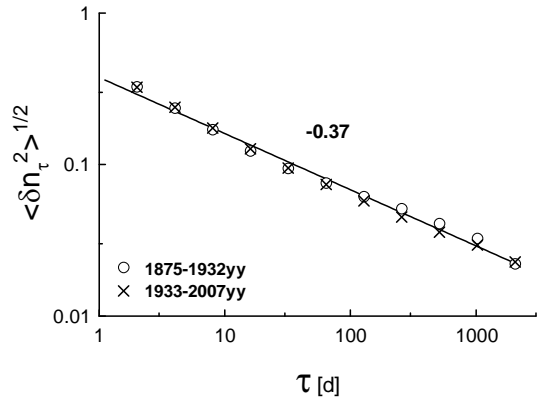} \vspace{-5cm} \caption{
The standard deviation for $\delta n_{\tau}$ vs. $\tau$ (in log-log scales) 
for the daily aa-index \cite{wdc}: 
circles correspond to the Historic period and crosses correspond to the Modern period. }
\end{figure}

At the Earth itself the solar wind induced activity is measured by geomagnetic indexes such as aa-index 
(in units of 1 nT). This, the most widely used long-term geomagnetic index \cite{may}, presents long-term 
geomagnetic activity and it is produced using two observatories at nearly antipodal positions on 
the Earth's surface. The index is computed from the weighted average of the amplitude of the 
field variations at the two sites. It was recently discovered that variations of this index 
are strongly correlated with the global temperature anomalies (see, for instance, \cite{cbf},\cite{land}). 
In this paper, however, we will be mostly interested in the clustering properties of the aa-index and 
their relation to the modulation produced by the photospheric convective motion. The point is that 
the data for the aa-index are available for both the Modern and the Historic periods. This 
allows us to check the suggestion that the Rayleigh number of the photospheric convection has the 
same order for the both mentioned periods. The transition between the two periods can be seen 
in Figure 6, where we show the cumulated aa-index $a(t)$
$$
S_{aa} (t) = \int_{0}^{t} a(t') dt'  \eqno{(2)}
$$
The arrow in this figure indicates beginning of the transitional solar cycle.

Figure 7 shows cluster-exponents for both the Modern and the Historic 
periods calculated for the low intensity levels of the aa-index (for the Historic period the level 
used in the calculations is aa-index=10nT, whereas for the Modern period the level is aa-index=25nT, 
the aa-index was taken from \cite{wdc}). 
The low levels of intensity were chosen in order to avoid effect of the extreme phenomena (magnetic 
storms and etc.). The straight line in Fig. 7 is drawn to show the expected value of the cluster-exponent 
$\alpha \simeq 0.37$ for the both periods. It means that indeed for both the Historic and the Modern 
periods the Rayleigh number $Ra \sim 10^{11}$ (cf. Fig. 4) in the solar photosphere.

\section{Impact of the solar dynamics on Earth climate}

\begin{figure} \vspace{-0.5cm}\centering
\epsfig{width=.45\textwidth,file=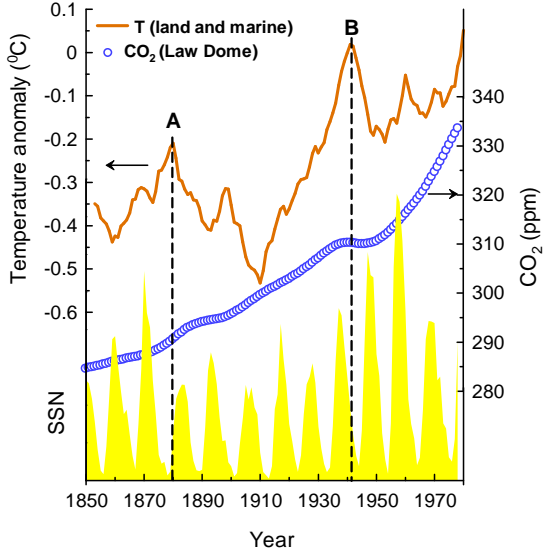} \vspace{-3cm}
\caption{Global temperature anomaly (solid curve, combined land and
marine) \cite{temp} and atmospheric $CO_2$ (circles) \cite{lawdome}
vs time. Relevant daily SSN are also shown. }
\end{figure}
\begin{figure} \vspace{-0.5cm}\centering
\epsfig{width=.45\textwidth,file=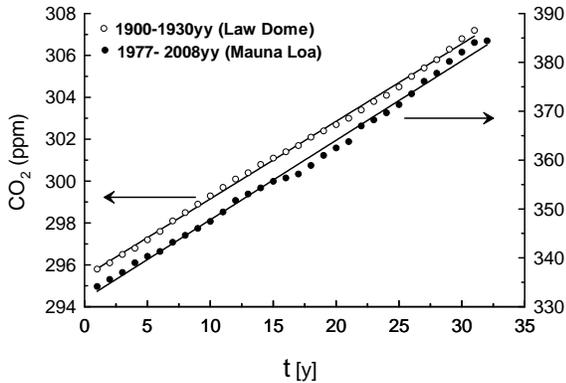} \vspace{-3cm}
\caption{Comparison of the $CO_2$ growth for the periods 1900-1930yy
\cite{lawdome} and 1977-2008yy \cite{maunaloa}. The solid straight
lines (best fit) indicate a linear growth.}
\end{figure}
\begin{figure} \vspace{-0.5cm}\centering
\epsfig{width=.45\textwidth,file=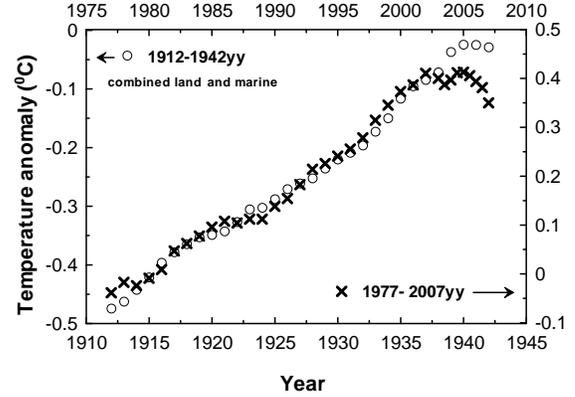} \vspace{-4cm} \caption{
Comparison of the temperature anomaly (combined land and marine)
\cite{temp} for the periods 1912-1942yy and 1977-2007yy. }
\end{figure}

The transitions of the solar convection zone dynamics can affect the earth climate 
through (at least) two channels. 
First channel is a direct change in the heat and
light output of the Sun, especially during the transitional solar cycle 1933-1944yy. 
Second channel is related to the strong increase of the magnetic field output in the
interplanetary space through the sunspots. The interplanetary
magnetic field interacts with the cosmic rays. Therefore, the change
in the magnetic field intensity can affect the Earth climate through
the change of the cosmic rays intensity and composition (see, for instance, 
Refs. \cite{uk},\cite{shav},\cite{kirk}). Let us look more closely 
to the transition B in Fig. 1. The Fig. 8
shows global temperature anomaly (the solid curve) \cite{temp} and
atmospheric $CO_2$ mixing
ratios (the circles) \cite{lawdome} vs. time. Relevant
SSNs are also shown. The dashed straight line (corresponding to the peak B) separates between
Historic and Modern periods. One can see, that the transitional
solar cycle (1933-1944yy) is characterized by dramatic changes both
in the global temperature (a huge peak) and in the atmospheric
$CO_2$ (complete suppression of the growth). After the transitional
period one can observe (during the three solar cycles: 1944-1977yy) certain
growth in the temperature anomaly and an unusually fast growth in
the atmospheric $CO_2$. These three solar cycles seems to be an
aftershock adaptation of the global climate to the new conditions.
Then, starting from 1977 year, the atmospheric $CO_2$ growth returns
to its linear trend as before the transition B, but now the {\it
rate} of the growth is about {\it four} times larger than before the
transition (see Fig. 9). On the other hand,  it can be seen from Fig. 1 
that the temperature anomaly before the maximum C 
returns to about the same growth pattern as it was just before the transition B. 
A quantitative comparison of these patterns, shown in Figure 10, indicates only about 
20\% difference in the growth rate (cf. also  Refs. \cite{ah}, \cite{gt}). 
The thirty year periods used for this comparison 
could find a support in the Appendix B.\\

If one consider the impact mechanism, related to the energetic cosmic (charged) 
particles and the corresponding screen effect of the magnetic fields, one could expect 
that this impact mechanism should be also sensitive to the 11-year solar cycle 
variability (though considerably less than to the transitional effects). To find 
fingerprints of this variability in the global temperature anomaly data is not a trivial 
task due to comparatively (to the 11-year cycle) short period of observations and due to 
the statistically {\it non}-stationary character of these data (with considerable time 
trends). To extract these fingerprints we will use a time series decomposition method 
developed in \cite{ks},\cite{bd},\cite{k}. A time series is decomposed into a trend (like that shown in Fig. 1 
as the solid curve) and a fluctuating (noise-like) components (see Appendix A). 
The method applies state space modeling and 
Kalman filter. Parameters are estimated by maximum likelihood method. Obtained by this way 
the fluctuating (noise-like) component of the (filtered) signal presents a statistically 
stationary time series and, therefore, it can be subject of a spectral analysis. Result of this 
analysis for the global temperature anomaly (combined land and marine \cite{temp}) is shown in Figure 11. 
Although we have deal with a comparatively 
short data series one can clear see the main spectral peak corresponding to the (about) 11-year 
{\it solar} cycle impact (cf. Refs. \cite{shav},\cite{swest},\cite{swest2}).  If one also introduces an  
autoregressive (AR) component into the time series decomposition (see Appendix A) to describe the fluctuating part 
of the global temperature anomaly, then one can compute energy spectrum of the AR component. Such spectrum 
is shown in figure 12. One can clear see the spectral peak corresponding to the solar 11-year cycle impact. 

\begin{figure} \vspace{-0.5cm}\centering
\epsfig{width=.45\textwidth,file=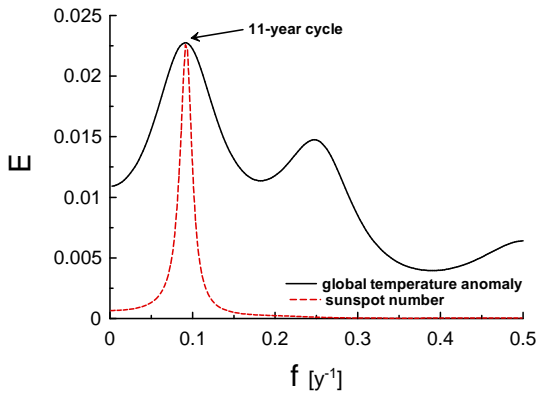} \vspace{-5cm}
\caption{The solid curve corresponds to energy spectrum of the fluctuating 
(noise-like) component 
of the decomposed global temperature anomaly time series (combined land and
marine \cite{temp}). The dashed curve corresponds to energy spectrum of fluctuating component of the SSN data.}
\end{figure}
\begin{figure} \vspace{-0.7cm}\centering
\epsfig{width=.45\textwidth,file=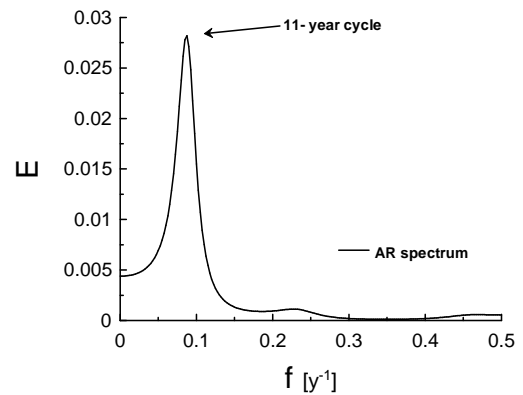} \vspace{-4.7cm}
\caption{Energy spectrum of the AR (fluctuating) component 
of the decomposed global temperature anomaly time series (combined land and
marine \cite{temp}). }
\end{figure}
\begin{figure} \vspace{-0.5cm}\centering
\epsfig{width=.45\textwidth,file=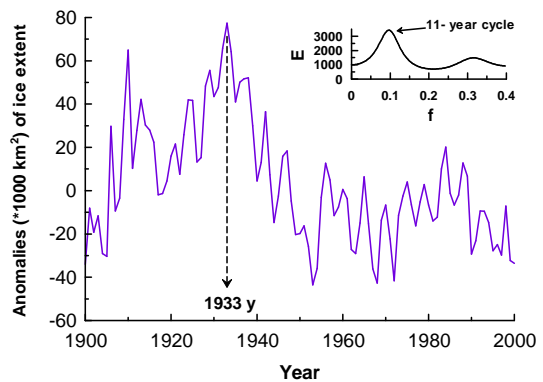} \vspace{-4.5cm}
\caption{Trend of anomalies of ice August extent in the Chukchi sea \cite{pol2}. The insert 
shows spectrum of fluctuating (noise-like) component of the time series decomposition.} 
\end{figure}
It is also interesting to look at the reduction in arctic ice extent. Figure 13 
shows a trend component of the anomalies of ice August extent in the 
Chukchi sea (the data are taken from \cite{pol2}) obtained by the time series decomposition. 
The dashed straight line indicates beginning of the {\it transitional} solar cycle. 
Insert to Fig. 13 shows spectrum of corresponding fluctuating (noise-like) component of the 
(filtered) signal (cf. Fig. 11). The dashed straight line indicates beginning of the {\it transitional} 
solar cycle. This arctic sea is located sufficiently far from the 
North Atlantic to diminish its influence, whereas influence of the Pacific ocean on the Chukchi sea 
local climate is not very significant.

\section{A forecast}

At present time we can already observe a peak C in the global temperature anomaly (see Fig. 1). 
The last observation can be 
considered as an indication of the current transition to a new section IV in 
the solar activity. The fact that in the 'last' period (1977-2007yy) the
temperature anomaly growth returned to the same pattern as just
before the transition peak B (see Fig. 10) provides an additional support 
to this suggestion. 

\begin{figure} \vspace{-1cm}\centering
\epsfig{width=.45\textwidth,file=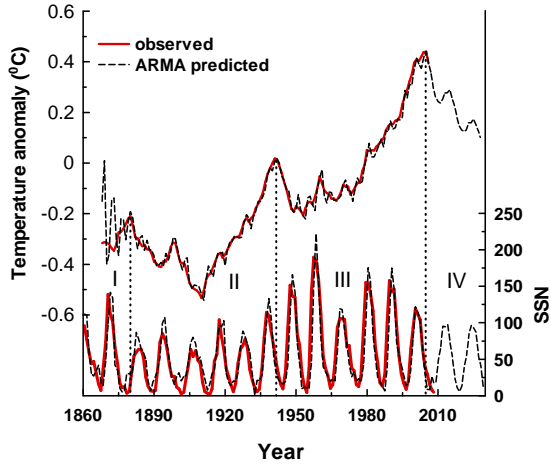} \vspace{-4cm}
\caption{The upper solid curve corresponds to the observed yearly global temperature anomaly data 
(7-year smoothed) \cite{temp}. 
The lower solid curve corresponds to the observed SSN data (year-smoothed) \cite{belg}.
The dashed curves are an ARMA-prediction: for SSN and for global temperature anomaly. }
\end{figure}
\begin{figure} \vspace{-0.5cm}\centering
\epsfig{width=.45\textwidth,file=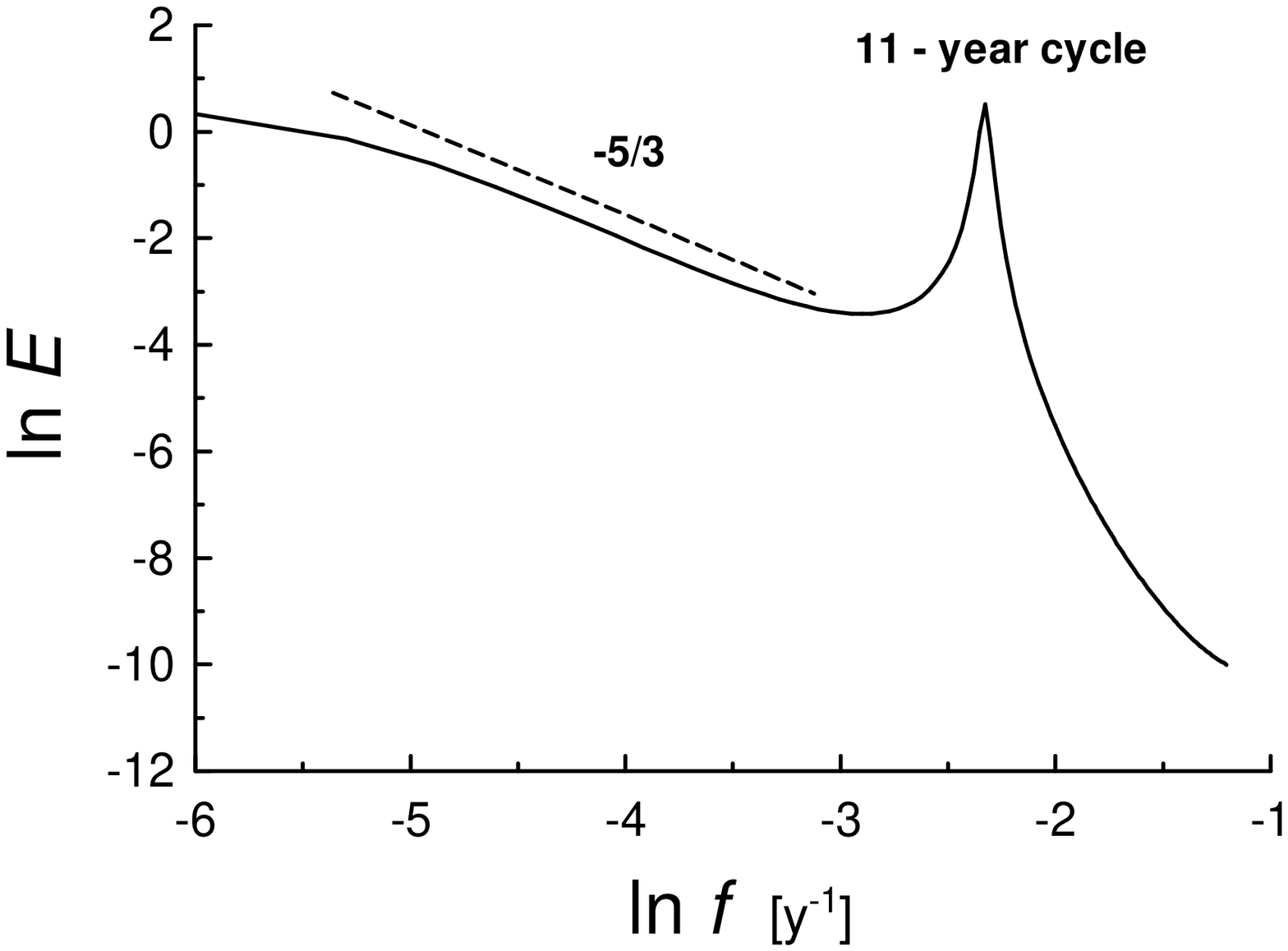} \vspace{-4.5cm}
\caption{ARMA spectrum of the global temperature anomaly fluctuations in the ln-ln scales. 
The straight dashed line is drawn to indicate 
the Kolmogorov scaling law $E(f) \sim f^{(-5/3)}$.}
\end{figure}
\begin{figure} \vspace{-0.5cm}\centering
\epsfig{width=.45\textwidth,file=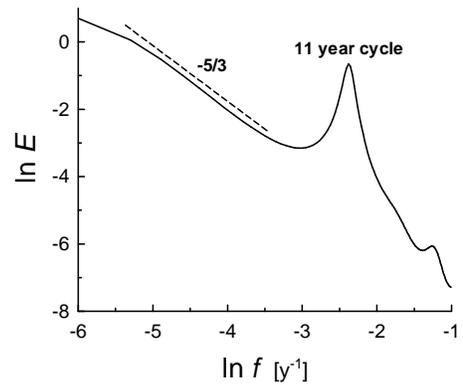} \vspace{-4.5cm}
\caption{Spectrum of the galactic cosmic ray count rate fluctuations in the ln-ln scales (the reconstructed 
data for period 1611-2007yy have been taken from \cite{usos-recon}). The straight dashed line 
is drawn to indicate the Kolmogorov scaling law $E(f) \sim f^{(-5/3)}$.}
\end{figure}
To make a forecast for the new section IV let us assume that there was a storage of certain part of 
the energy coming from the radiative 
zone during the period of the section II (Fig. 1). Then, in the period of the section III 
this stored energy was released and transported to the Sun surface.  
This release was not in the form of the thermal convection, but in the form of strongly concentrated toroidal 
magnetic field structures, which results in the intense emergence of the sunspots in the active parts of the 
Modern period (section III in Fig. 1). It is interesting that the storage and release periods 
have about the same length: $\simeq 60$ years (see also Appendix B). The storage of energy in a growing
magnetic field at the bottom of the convection zone or in the overshoot (by the convective downflows) 
area just under the bottom can be considered as the most plausible for this scenario. 
Convective overshooting provides means to store magnetic energy below the convection zone by 
resisting the buoyancy effect for large time scales. Level of subadiabaticity in the overshoot layer, 
needed in order to store the magnetic flux for sufficiently long times, can be significant. Expected 
intensity of $10^5~G$ results an energy density that is in order of magnitude larger than equipartition value. Though, 
suppression of convective motion by the magnetic flux tubes themselves can be considered as 
an additional factor allowing to achieve the sufficient level of subadiabaticity (see, for instance 
\cite{sw},\cite{brand1},\cite{th}). 

 Since the section III in the solar activity was a release one, it is naturally to assume that the next 
section (IV) in the solar activity will be a 'storage' one, similar to the 'storage' section II.  As for the corresponding Earth temperature anomaly, the section IV is also expected to be similar to the section II. But the temperature anomaly from the beginning of the section IV will be shifted 
upward on about $0.6^oC$ in comparison with the section II. It then will be 
not surprising if the total length of the section IV also will be about 60 years (cf. Appendix B) and the section will be 
finished with a maximum {\bf D} higher than $0.6^oC$.  A very schematic presentation for a first part 
of the section IV is shown in figure 14 (see Appendix A for an Auto-Regressive-Moving-Average (ARMA) model used for the prediction, cf. also Ref. \cite{panel}).  Figure 15 shows corresponding ARMA model spectrum of the global temperature anomaly fluctuations (in ln-ln scales). A dashed 
straight line in this figure indicates a scaling with '-5/3' exponent: $E(f) \sim f^{-5/3}$. 
Although, the scaling interval is short, the value of the exponent is rather intriguing. 
This exponent is well known in the theory of fluid (plasma) turbulence and corresponds to so-called 
Kolmogorov's cascade process (see, for instance \cite{my}). This process is very universal for turbulent 
fluids and plasmas \cite{gibson},\cite{clv}. For turbulent processes in Earth and in Heliosphere 
the Kolmogorov-like spectra with such large time scales cannot exist. Therefore, one should think about 
a Galactic origin of Kolmogorov turbulence (or turbulence-like processes \cite{gibson1})
with such large time scales. This is not surprising if we recall possible role of the galactic cosmic rays for Earth climate (see, for instance, \cite{uk},\cite{shav},\cite{kirk}). In order to support this point 
we show in figure 16 spectrum of galactic cosmic ray intensity at the Earth's orbit (reconstruction for period 1611-2007yy \cite{usos-recon}, cf. also Ref. \cite{b1}). One can compare Fig. 16 with Figs. 15 corresponding to the global temperature anomaly fluctuations (see also Appendix C). Presence of the turbulent-like component in the low-frequency fluctuations makes any long-range prediction a very difficult task (see also Appendix B on a chaotic element in solar dynamics). A chaotic element in global climate dynamics itself (Appendix C) presents another problem for the forecast. Such internal forcing agents as volcanoes and anthropogenic greenhouse gases, for instance, have a potential to change drastically the future dynamics of the global climate. \\ 

\begin{figure} \vspace{-1cm}\centering
\epsfig{width=.45\textwidth,file=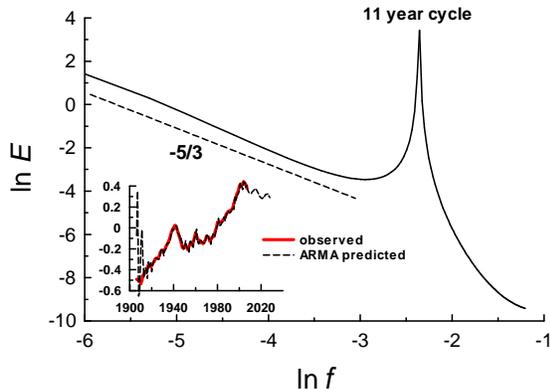} \vspace{-4.5cm}
\caption{Spectrum of the global temperature anomaly fluctuations in the ln-ln scales for a 
cutted observed base ARMA model. The straight dashed line is drawn to indicate 
the Kolmogorov scaling law $E(f) \sim f^{(-5/3)}$.}
\end{figure}

The author is grateful to J.J. Niemela, to K.R. Sreenivasan, 
and to SIDC-team, World Data Center for the Sunspot Index,
Royal Observatory of Belgium for sharing their data and discussions. 
The author is also grateful to C.H. Gibson and to R.A. Treumann for their help in improving the paper. 
A software provided by K. Yoshioka was used at the model computations. 

\section{Appendix A}

The above applied time series decomposition method was developed in the Refs. \cite{ks},\cite{bd},\cite{k}. 
In this method a nonstationary mean time series is decomposed into trend, autoregressive and 
noise components. A random process  $X_t$ in discrete time $t$ is defined by 
the following expression: 
$$
X_t=\tau_t+ \varepsilon_t  + AR_t^{(m)}\eqno{(A1)}
$$
where $\tau_t$ is a trend component, $\varepsilon_t$ is an observational noise component, and 
$$
AR_t^{(m)}=\sum_{i=1}^{m} a_i X_{t-i} + \eta_t^{(1)}  \eqno{(A2)}
$$ 
is a globally stationary autoregressive (AR) component of order $m$ ($a_i$ are the coefficients of a 
recursive filter, and $\eta_t^{(1)}$ is a Gaussian white noise).  

A $k$-th order stochastically perturbed difference equation describes evolution of the trend component 
$$
\Delta^k \tau_t =\eta_t^{(2)}   \eqno{(A3)}
$$
where $\eta_t^{(2)}$ is a Gaussian white noise. 

In the Ref. \cite{k} algorithms, using state space modeling and Kalman filter, are suggested for 
estimation of the model parameters by the maximum likelihood method. 

Using so-called autoregressive moving average (ARMA) model one could try to make a prediction. 
For this model a random process  $X_t$ in discrete time $t$ is defined by the following expression:  
$$
X_t=\varepsilon_t  + AR_t^{(m)}+ \sum_{i=1}^{n} b_i \varepsilon_{t-i} \eqno{(A4)}
$$
where $\varepsilon_t$ is Gaussian white noise with zero mean \cite{ks},\cite{bd},\cite{k}. Again, state space modeling and Kalman filter, are used  for estimation of the model parameters by the maximum likelihood method. Figure 14
shows an optimistic prediction produced with this model for a first part of the section IV. Though, continuation 
of the analogy to the section II makes the scenario much less optimistic. A quasi-Newton method and dynamic restrictions were used for optimization of the model parameters. 

We can also play with the model by cutting the {\it observed} time series. This procedure reduces relative 
contribution of the low-frequency fluctuations, presumably of Galactic origin (cf. Fig. 17 and Fig. 15). Then, 
the insert to the Fig. 17 shows less steep decline of the global temperature anomaly at this (cutted observed base) version of the ARMA model prediction. This game has also another end. One can assume that using a more {\it long} 
time series (observed base), than that we have used in computing the results shown in Fig. 14, one could 
obtain even more optimistic prediction for the first part of the section IV.    

\section{Appendix B}

The long-range reconstructions of the sunspot number fluctuations 
(see, for instance, Refs. \cite{sol},\cite{usos1}) allow us to look on the solar transitional 
dynamics from a more general point of view. In figure 18 we show a spectrum of such 
reconstruction for the last 11,000 years (the data, used for computation of the spectrum, 
is available at \cite{usos-recon}). The spectrum was computed using the autoregressive model 
(with maximum entropy method \cite{k}). A semi-logarithmical representation was used in 
the figure to show an exponential law 
$$
E(f) \sim e^{-f/f_e}   \eqno{(B1)}
$$
The straight line is drawn in Fig. 18 to indicate the exponential law Eq. (B1). Slope of the 
straight line provides us with the characteristic time scale $T_e=1/f_e\simeq 176 \pm 7$y. 
The exponential decay of the spectrum excludes the possibility of random behavior and indicates the {\it chaotic} 
behavior of the time series \cite{sig},\cite{sun}. It is well known that low-order dynamic (deterministic) systems
have as a rule exponential decay of $E(f)$ (see, for instance, \cite{sun},\cite{o}). As for infinite dimensional dynamic systems 
with chaotic attractors it is interesting to compare Fig. 18  with figure 3 of the Ref. \cite{fa}. 
It should be noted that the 176y 
period is the third doubling of the period 22y. The 22y period corresponds to the Sun's magnetic 
poles polarity switching (see also Appendix C).
\begin{figure} \vspace{-1cm}\centering
\epsfig{width=.45\textwidth,file=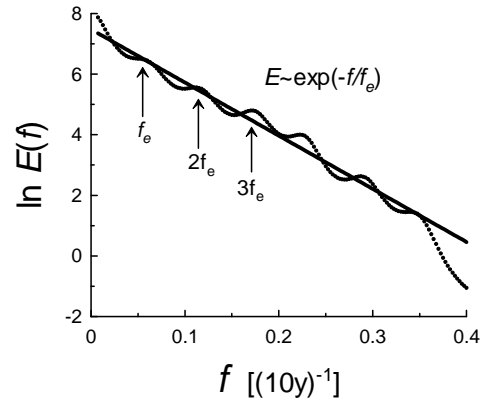} \vspace{-4.5cm}
\caption{Spectrum of the sunspots number fluctuations in the ln-linear scales (the reconstructed data 
for the last 11,000 years have been taken from \cite{usos-recon}). The straight line is drawn to indicate 
the exponential law Eq. (B1).}
\end{figure}

The exponential spectrum can be also 
produced by a series of Lorentzian pulses with the average width of the individual pulses equals to $\tau$ 
(though, the distribution of widths of the pulses should be fairly narrow to result in the 
exponential spectrum).

In Fig. 18 a local maximum corresponding to 
the frequency $f_e$ and its first harmonics have been indicated by arrows. It should be noted 
that the harmonic $2f_e$ corresponds to the well known period $T\simeq 88y$ (see, for instance, Ref. \cite{fg}), 
and the harmonic $3f_e$ corresponds to the period $T \simeq 60y$ (cf. section "A FORECAST").  
 
\section{Appendix C}
 
\begin{figure} \vspace{-0.5cm}\centering
\epsfig{width=.45\textwidth,file=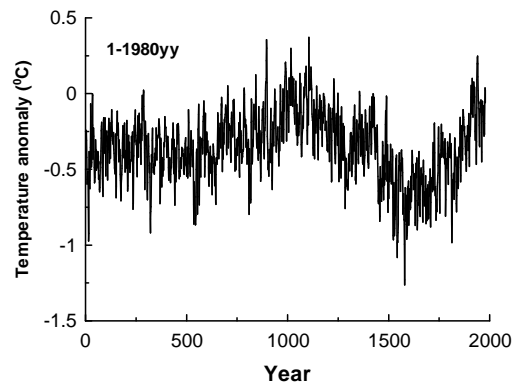} \vspace{-4.5cm}
\caption{A reconstruction of Northern Hemisphere temperature
anomaly for the past 2,000 years (the data was taken
from Ref. \cite{paleoT}).}
\end{figure}
It is also useful to consider the above results in content
of multicentennial and millennial timescales. Figure 19
shows a reconstruction of Northern Hemisphere temperatures
for the past 2,000 years (the data for this figure were
taken from Ref. \cite{paleoT}). This multi-proxy reconstruction
was performed by the authors of Ref. \cite{moberg} using combination
of low-resolution proxies (lake and ocean sediments)
with comparatively high-resolution tree-ring data.
Figure 20 shows a power spectrum of the data set calculated
using the maximum entropy method (as for the
spectra shown in Figs. 16 and 18) in the frames of the autoregressive
model, because it provides an optimal spectral
resolution even for small data sets (see also \cite{o}).
The spectrum exhibits a rather wide peak indicating a
periodic component with a period around 22 y, and a
broad-band part with exponential decay. A semilogarithmical
plot was used in Fig. 20 (cf Fig. 18) in order to
show the exponential decay more clearly (at this plot the
exponential decay corresponds to a straight line). Both
stochastic and deterministic processes can result in the
broad-band part of the spectrum, but the decay in the
spectral power is different for the two cases. The exponential
decay indicates that the broad-band spectrum
for these data arises from a deterministic rather than a
stochastic process. Indeed, for a wide class of deterministic
systems a broad-band spectrum with exponential
decay is a generic feature of their chaotic solutions (see,
for instance, Refs. \cite{sig},\cite{o},\cite{fa} and Appendix B).

\begin{figure} \vspace{-1cm}\centering
\epsfig{width=.45\textwidth,file=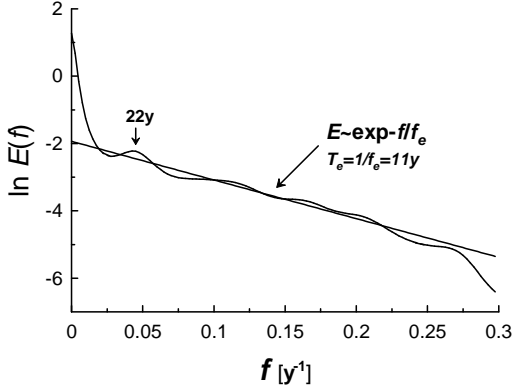} \vspace{-4.5cm}
\caption{Spectrum of the data, shown in Fig. 19, in semilogarithmical
scales. The straight line indicates the exponential
decay Eq. (B1).}
\end{figure}
\begin{figure} \vspace{-0.3cm}\centering
\epsfig{width=.45\textwidth,file=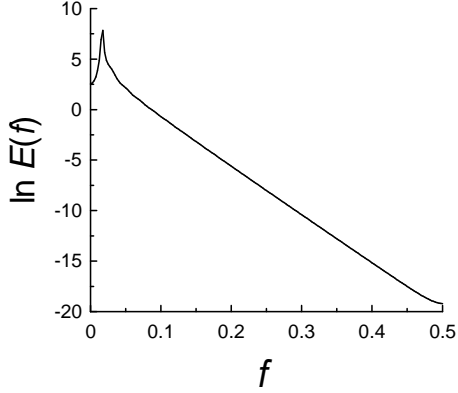} \vspace{-5cm}
\caption{Spectrum of the chaotic fluctuations of the x-
component for the R\"{o}ssler system ($a = 0.15$, $b = 0.20$, $c =
10.0.$)}
\end{figure}
\begin{figure} \vspace{-1cm}\centering
\epsfig{width=.45\textwidth,file=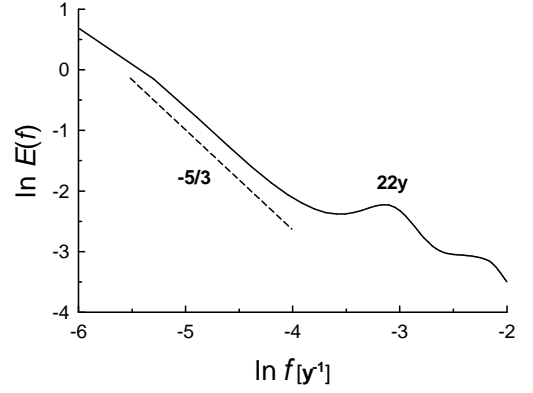} \vspace{-4.5cm}
\caption{Spectrum of the data, shown in Fig. 19, in ln-ln
scales. The dashed straight line indicates the Kolmogorov-like
spectrum: $E(f)\sim f^{-5/3}$. The high-frequency part has
been cutted in order to show the low-frequency part.}
\end{figure}

In order to illustrate this and another significant feature
of the chaotic power spectra we show in figure 21 a
power spectrum for the R\"{o}ssler system \cite{ros}
$$
\frac{dx}{dt} = -(y + z);~~  \frac{dy}{dt} = x + a y;~~ 
\frac{dz}{dt} = b + x z - c z \eqno{(C1)}
$$ 
chaotic solution, where $a$, $b$ and $c$ are parameters.
In this figure one can see a typical picture: a narrow-band
peak (corresponding to the fundamental frequency
of the system) in a low-frequency part and a broad-band
exponential decay in a high-frequency part of the spectrum.
Nature of the exponential decay of the power spectra
of the chaotic systems is still an unsolved mathematical
problem. A progress in solution of this problem
has been achieved by the use of the analytical continuation
of the equations in the complex domain (see, for 
instance, \cite{fu},\cite{fm}). In this approach the exponential decay
of chaotic spectrum is related to a singularity in the
plane of complex time, which lies nearest to the real axis.
Distance between this singularity and the real axis determines
the rate of the exponential decay. If parameters of
the dynamical system periodically fluctuate around their
mean values, then at certain (critical) intensity of these
fluctuations an additional singularity (nearest to the real
time axis) can appear. Distance between this singularity
and the real axis is determined by period of the periodic
fluctuation of the system's parameters. Therefore, exponential
decay rate of the broad-band part of the system
spectrum equals the period of the parametric forcing.
The chaotic spectrum provides two different characteristic
time-scales for the system: a period corresponding to
fundamental frequency of the system, $T_{fun}$, and a period
corresponding to the exponential decay rate, $T_e = 1/f_e$
(cf Eq. (B1)). The fundamental period $T_{fun}$ can be estimated
using position of the low-frequency peak, while the
exponential decay rate period $T_e = 1/f_e$ can be estimated
using the slope of the straight line of the broad-band part
of the spectrum in the semilogarithmical representation
(Figs. 20 and 21). From Fig. 20 we obtain $T_{fun} \simeq 22\pm 2$y
and $T_e \simeq 11 \pm 1y$ (the estimated errors are statistical
ones). Thus, the solar activity (SSN) period of 11 years
is still a dominating factor in the chaotic temperature
fluctuations at the millennial time scales , although it is
hidden for linear interpretation of the power spectrum (cf
Ref. \cite{b3}). In the nonlinear interpretation the additional
period $T_{fun} \simeq 22y$ might correspond to the fundamental
frequency of the underlying nonlinear dynamical system.
It is surprising that this period is close to the 22y period
of the Sun's magnetic poles polarity switching (cf
Appendix B). It should be noted that the authors of Ref. \cite{mu} found a persistent 
22y cyclicity in sunspot activity, presumably related to 
interaction between the 22y period of magnetic poles polarity switching 
and a relic solar (dipole) magnetic field. Therefore, one cannot rule out 
a possibility that the broad peak, in a vicinity of frequency corresponding to the 22y period, 
is a quasi{\it-linear} response of the global temperature to the weak periodic 
modulation by the 22y cyclicity in sunspot activity. I.e. strong enough 
periodic forcing results in the non-linear (chaotic) response whereas a weak periodic forcing 
results in a quasi-periodic response.
  
Finally, figure 22 shows the same power spectrum as in 
Fig. 20 but in ln-ln scales, where the dashed straight
line indicates the Kolmogorov-like behavior: $E(f) \sim
f^{-5/3}$, in the lower-frequency part of the spectrum (see
section 'A Forecast' and Fig. 16 for relation between this
type of spectrum and galactic turbulence).

\end{document}